\documentclass[3p,twocolumn]{elsarticle}
\usepackage{graphicx,color,mathrsfs,amsmath,amssymb,amsthm,amsopn,bm}
\usepackage[nottoc]{tocbibind}
\usepackage[normalem]{ulem}
\newlength{\lslashl}
\def\FMslash#1{\settowidth{\lslashl}{$#1$}\makebox[\lslashl]{\makebox[0mm]{$/$}\makebox[0mm]{$#1$}}}
\def\qslash{\FMslash q}

\def\back{\negthickspace\negthickspace\negthickspace\negthickspace\negthickspace\negthickspace\negthickspace\negthickspace\negthickspace\negthickspace\negthickspace\negthickspace}
\usepackage[normalem]{ulem} 

\begin{document}

\begin{frontmatter}

\title{Medium Effects in $\rho$-Meson Photoproduction}

\author[TAMU]{F. Riek}
\ead{friek@comp.tamu.edu}
\author[TAMU]{R. Rapp}
\ead{rapp@comp.tamu.edu}
\author[Argonne]{T.-S. H. Lee}
\ead{lee@phy.anl.gov}
\author[TAMU]{Yongseok Oh}
\ead{yoh@comp.tamu.edu}
\address[TAMU]{Cyclotron Institute and Physics Department,
Texas A\&M University, College Station, Texas 77843-3366, USA}
\address[Argonne]{Physics Division, Argonne National Laboratory, Argonne, Illinois 60439, USA}

\date{\today}

\begin{abstract}
We compute dilepton invariant mass spectra from the decays of $\rho$ mesons
produced by photon reactions off nuclei.
Our calculations employ a realistic model for the $\rho$ photoproduction
amplitude on the nucleon which provides fair agreement with measured
cross sections.
Medium effects are implemented via an earlier constructed $\rho$
propagator based on hadronic many-body theory.
At incoming photon energies of 1.5--3~GeV as used by the CLAS experiment
at JLAB, the average density probed for iron targets is estimated at about
half saturation density.
At the pertinent 3-momenta the predicted medium effects on
the $\rho$ propagator are rather moderate.
The resulting dilepton spectra approximately agree with recent CLAS data.
\end{abstract}

\begin{keyword}
photoproduction, $\rho$ meson, in-medium properties
\PACS{21.65.Jk, 25.20.Lj, 14.40.Cs}
\end{keyword}

\end{frontmatter}

\section{Introduction}
The investigation of hadron properties in hot and/or dense matter is
of fundamental interest in the context of approaching the transition(s)
into a chirally restored and/or deconfined plasma of quarks and gluons.
Intriguing effects have been observed in dilepton spectra measured in
high-energy heavy-ion collisions~\cite{Arnaldi:2006jq,Adamova:2006nu},
which are consistent with a strong broadening of the $\rho$-meson spectral
function by about a factor of $\sim$3 in hot and dense hadronic
matter~\cite{Rapp:1999ej,vanHees:2007th}.
The medium modifications of the $\rho$ are believed to be largely driven
by the baryonic component of the medium.
The rapid expansion of the fireball formed in heavy-ion reactions implies
that the emission spectra encode a rather large range of temperatures and
densities of the evolving medium.
It is therefore desirable to test the medium effects in a static environment,
such as provided by ground-state nuclei.
Hadronic models predict appreciable medium effects in cold nuclear matter,
e.g., an increase of the width of low-momentum $\rho$ mesons at saturation
density by a factor of 2--3 over its vacuum value~\cite{Rapp:1999ej}.
Several experiments have recently been conducted to measure $\rho$
production off nuclei, in both proton-~\cite{Naruki:2005kd} and
photon-induced~\cite{Huber:2003pu,Clas:2007mga,Wood:2008ee} reactions.
In Ref.~\cite{Naruki:2005kd} a rather small dilepton signal for the
$\rho$ has been reported with a mass distribution compatible with a
dropping mass, while Refs.~\cite{Huber:2003pu} and \cite{Wood:2008ee}
extracted a moderate broadening with little, if any, mass shift in
$\pi^+\pi^-$ and $e^+e^-$ mass spectra, respectively.
An inherent feature of nuclear production experiments is that rather large
projectile energies are required to supply the rest mass of the $\rho$.
These impart an appreciable 3-momentum on the $\rho$ meson relative
to the nucleus which enhances the probability for decays outside
the nucleus thus reducing the effective density probed by these
experiments.
Nevertheless, valuable constraints for cold nuclear matter effects and
their 3-momentum dependence on existing models for in-medium
$\rho$ spectral functions can be expected.

As in heavy-ion reactions, dileptons are of special interest due
to their negligible final-state interactions.
However, the initial states in heavy-ion collisions and nuclear production
experiments are quite different.
In the former case, the simplifying assumption of a thermal heat bath can be
made, while in the latter case a reliable description of the elementary
production process is mandatory.
The pertinent baseline reaction on a single nucleon,
$\gamma\,N\,\rightarrow\,e^+\,e^-\,N$, has been studied in several theoretical
works~\cite{Schafer:1994vr,Friman:1995qm,Effenberger:1999ay,Oh:2003aw,Lutz:2005yv}.
Generally, the low-energy cross section is dominated by baryon resonance 
formation,
while at photon energies of $\sim$1.5--2~GeV $t$-channel exchange processes
are expected to take over. Applications to nuclear targets can be found
in Ref.~\cite{Effenberger:1999ay}, where a schematic model for the production
process has been implemented into a transport simulation for final-state
interactions, as well as in Ref.~\cite{Wood:2008ee} for the CLAS 
data~\cite{Clas:2007mga,Wood:2008ee}.

In the present work we combine a microscopic model for $\rho$ photoproduction
on the nucleon~\cite{Oh:2003aw} with an in-medium $\rho$ spectral function
computed in hadronic many-body theory~\cite{Rapp:1997ei,Rapp:1999us}.
The production model is largely based on meson/Pomeron exchange which
properly accounts for the cross section above photon energies of 2~GeV.
At lower energies, we supplement additional $s$-channel resonance excitations
with parameters directly taken from the in-medium selfenergy of the $\rho$
spectral function~\cite{Rapp:1999ej,Rapp:1999us}.
This establishes consistency between the production process and in-medium
effects, and leaves no additional free parameters for the resulting 
cross sections for $\rho$ production and dilepton invariant-mass spectra.

In Sec.~\ref{sec_rho-prod} the $\rho$ photoproduction model on the nucleon
is presented and checked against total cross sections and dilepton
invariant-mass spectra for deuteron targets. In Sec.~\ref{sec_med} we apply
our model to dilepton spectra off nuclei utilizing the in-medium $\rho$
spectral function at densities estimated from the decay kinematics
corresponding to the incoming photon spectrum in the CLAS experiment.
We finish with conclusions in Sec.~\ref{sec_concl}.


\section{$\rho$ Photoproduction on the Nucleon}
\label{sec_rho-prod}

\begin{figure}[!t]
\includegraphics[scale=0.85]{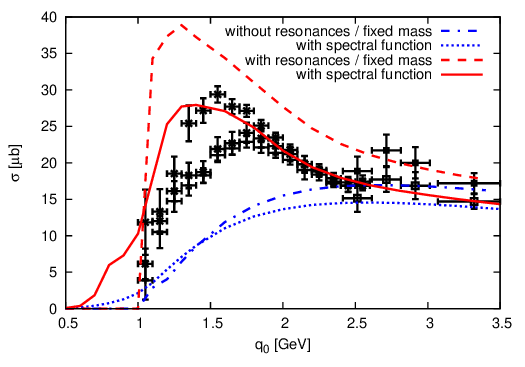}
\caption{Total cross section for $\gamma\,p\,\rightarrow\,p\,\rho^0$ as 
a function of incident photon energy $q_0^{}$ in the laboratory frame.
The calculations are based on either a fixed $\rho$-mass of 770~MeV
(dash-dotted and dashed line) or a full vacuum spectral function
(dotted and solid line), either with (dashed and solid line) or without
(dash-dotted and dotted line) baryon resonances.
Data are from Refs.~\cite{Wu:2005wf,ABBHHM:1968ke}.}
\label{fig_xsec-tot}
\end{figure}

Our starting point is the photoproduction amplitude for
$\gamma\,p\,\rightarrow\,e^+\,e^-\,p$ developed by two of
us~\cite{Oh:2003aw}. It accounts for $\sigma$, $f_2$, $2\pi$ and Pomeron
$t$-channel exchange as well as nucleon $s$- and $u$-channel pole
contributions, and gives a good description of experimental cross
sections at photon energies above $\sim$2~GeV, cf. the dotted line in
Fig.~\ref{fig_xsec-tot}.
(A similar model in Ref.~\cite{Friman:1995qm} employs a stronger 
$\sigma$-exchange, see Ref.~\cite{Oh:2003aw,OK03} for further comparison.)
At smaller photon energies, as part of 
the photon beam used by CLAS
at JLAB ($q_0 \simeq 1 \mbox{--} 3.5$~GeV), baryon resonances are
expected to become important~\cite{Oh:2007jd}.
Here we adopt the same set of resonances as used in
Ref.~\cite{Rapp:1997ei} to describe total photoabsorption spectra
(to constrain the in-medium $\rho$ spectral function) summarized in
Tab.~\ref{tab_res}.\footnote{The $\rho NN$ formfactor has been reduced
to 600~MeV improving consistency with Ref.~\cite{Oh:2003aw}, together
with a 10\% reduction of the $\rho$-$N$-$N(1720)$ coupling constant. The
extra coupling to the $N$(2090) increases the $\rho$ production cross
section around photon energies of $\sim$2~GeV by $\sim$15\%.}
\begin{table}
\begin{center}
\begin{tabular}{c|c|c|c|c|c}
Resonance & $m_B$& $\Gamma_B^{\rm tot}$ & $\Gamma_{\rho\,N}$
& $\frac{f_{\rho N B}^2}{4\pi}$ & $\Lambda_B$\\
\hline
$\Delta(1232)$ & 1232 & 120 & N/A  & 16.2 & 700\\
$\Delta(1620)$ & 1620 & 145 & 35  & 2.1  & 700\\
$\Delta(1700)$ & 1700 & 300 & 110 & 2.5  & 1000\\
$\Delta(1905)$ & 1905 & 350 & 315 & 7.0  & 1200\\
$N(1440)$      & 1440 & 350 & 10  & 1.1  & 600\\
$N(1720)$      & 1720 & 200 & 100 & 4.16  & 600\\
$N(1520)$      & 1520 & 120 & 25  & 6.5  & 600\\
$N(2090)$      & 2090 & 414 & 150  & 1.0  & 1000
\end{tabular}
\caption{Resonance parameters (columns 2, 5 and 6) used in the elementary
photoproduction amplitude. The dimensionless coupling constants (all other
quantities are in [MeV]) are fixed
to reproduce the values for the total and partial vacuum on-shell
widths (columns 3 and 4) as in Refs.~\cite{Rapp:1999ej,Rapp:1999us} (the
cut-off values of Refs.~\cite{Rapp:1999ej,Rapp:1999us} are kept fixed).}
\label{tab_res}
\end{center}
\end{table}
Due to the rather large photon energies involved we employ relativistic
interaction vertices defined by the following $\rho$-$N$-$B$ Lagrangians:
\begin{eqnarray}
\mbox{}\hspace{-1cm} &&
\begin{array}{ll}
\mathcal{L}^{\frac{1}{2}\frac{1}{2}^+}=\frac{f_{\rho BN}}{m_{\rho}}\Bar{\Psi}_{R}\,\gamma_{5}\,\sigma^{\mu\nu}\,\tau_{i}^{}\rho^{i}_{\mu\nu}\,\Psi_{N}\,&+\,h.c.\,,\nonumber\\
\mathcal{L}^{\frac{1}{2}\frac{3}{2}^+}=\frac{f_{\rho BN}}{m_{\rho}}\Bar{\Psi}_{R}^{\mu}\,\gamma_{5}\,\gamma^{\nu}\,\tau_{i}^{}\,\rho^{i}_{\mu\nu}\,\Psi_{N}\,&+\,h.c.\,,\nonumber\\
\mathcal{L}^{\frac{1}{2}\frac{3}{2}^-}=\frac{f_{\rho BN}}{m_{\rho}}\Bar{\Psi}_{R}^{\mu}\,\gamma^{\nu}\,\tau_{i}^{}\,\rho^{i}_{\mu\nu}\,\Psi_{N}\,&+\,h.c.\,,\nonumber\\
\mathcal{L}^{\frac{3}{2}\frac{1}{2}^-}=\frac{f_{\rho BN}}{m_{\rho}}\Bar{\Psi}_{R}\,\sigma^{\mu\nu}\,T_{i}\,\rho^{i}_{\mu\nu}\,\Psi_{N}\,&+\,h.c.\,, \nonumber\\
\mathcal{L}^{\frac{3}{2}\frac{3}{2}^+}=\frac{f_{\rho BN}}{m_{\rho}}\Bar{\Psi}_{R}^{\mu}\,\gamma_{5}\,\gamma^{\nu}\,T_{i}\,\rho^{i}_{\mu\nu}\,\Psi_{N}\,&+\,h.c.\,,\nonumber\\
\mathcal{L}^{\frac{3}{2}\frac{3}{2}^-}=\frac{f_{\rho BN}}{m_{\rho}}\Bar{\Psi}_{R}^{\mu}\,\gamma^{\nu}\,T_{i}\,\rho^{i}_{\mu\nu}\,\Psi_{N}\,&+\,h.c.\,, \\ 
\end{array} \\
\mbox{}\hspace{-1cm} &&
\label{Lag}
\end{eqnarray}
where $\sigma^{\mu\nu}=\frac{i}{2}[\gamma^{\mu},\gamma^{\nu}]$, $\rho_{i}^{\mu\nu}=\partial^{\mu}\,\rho_{i}^{\nu}-\partial^{\nu}\,\rho_{i}^{\mu}$.
Isospin, spin and parity of the resonance $B$ are denoted by ${IJ^P}$,
and $\tau_{i}^{}$, $T_{i}$ are the usual isospin-1/2, $1/2 \to 3/2$ transition
matrices.\footnote{The spin-5/2 resonance is treated in a simplified way as
in Ref.~\cite{Rapp:1997ei} via a $\frac{3}{2}\frac{3}{2}^+$
state with amended spin factor.}
As in Ref.~\cite{Rapp:1997ei}, we utilize an improved version of the
vector dominance model (VDM)~\cite{Kroll:1967it,Friman:1997tc} which
allows for a direct $\gamma$-$N$-$B$ coupling and thus a better
simultaneous description of hadronic and radiative decay widths (the
$\gamma$-$N$-$B$ coupling follows by replacing $\rho_0^\mu$ with
$A^{\mu}$ in Eqs.~(\ref{Lag}); the parameter $r_B^{}=0.7$~\cite{Rapp:1997ei}
controls the deviation from naive VDM). The parameters (listed in
Tab.~\ref{tab_res}) are adjusted to recover the same partial decay width
for $B\to\rho N$ as the (updated) values in Ref.~\cite{Rapp:1999ej}
(based on Ref.~\cite{Rapp:1997ei}), including monopole formfactors
\begin{eqnarray}
F(|\vec{q}|)=\frac{\Lambda^2_{\rho BN}}{\Lambda^2_{\rho BN}+\vec{q}^{\,2}}
\label{Formfactor}
\end{eqnarray}
with cutoff parameters $\Lambda_{\rho BN}$~\cite{Rapp:1999ej}. The 
formfactors are consistently evaluated in the laboratory frame with $\vec{q}$ 
the three-momentum of the incoming photon (or $\rho$ in the nuclear rest frame).
The spin-1/2 and -3/2 baryon propagators are, respectively, taken
as
\begin{eqnarray}
&&\back S(q)=\frac{\qslash+m_B}{q^2-m_B^2+i\,m_B\,\Gamma_B}\, ,
\nonumber \\
&&\back S^{\mu\nu}(q)=\frac{\qslash+m_B}{q^2-m_B^2+i\,m_B\,\Gamma_B}\,
P^{\mu\nu}\, ,
\\
&&\back P^{\mu\nu}=g^{\mu\nu}-\frac{1}{3}\gamma^{\mu}\,
\gamma^{\nu}-\frac{2}{3}\frac{q^{\mu}q^{\nu}}{m_B^{2}}
+\frac{1}{3}\frac{q^{\mu}\gamma^{\nu}-q^{\nu}\gamma^{\mu}}{m_B} \, ,
\nonumber
\end{eqnarray}
with masses $m_B^{}$ and total widths $\Gamma_B$ as given in
Tab.~\ref{tab_res}. We furthermore assume a linear increase of the
in-medium resonance widths with density as in
Refs.~\cite{Rapp:1999ej,Rapp:1997ei}.
We have verified that neglecting the momentum dependence of $\Gamma_B$
in the propagators has an insignificant impact on our results.
Including $s$- and $u$-channel graphs, the baryon resonance parts
of the $\rho$-production amplitude take the form
\begin{eqnarray}
&&\back \mathcal{M}_{B}^{\mu\nu}=\chi_{I}^{}\,\frac{\mu_{B}^{}\,f_{\rho BN}}{2\,m_{\rho}}
\bar{u}(p^{\prime})\left[\Gamma_{VB}^{\mu}(k)\,S(p+q)\right.
\nonumber\\
&&\back \left. \quad \times\Gamma_{\gamma B}^{\nu}(q)
+\Gamma_{\gamma B}^{\nu}(q)\, S(p-k)\,\Gamma_{VB}^{\mu}(k)\right]u(p),
\nonumber\\
&&\back \mathcal{M}_{B}^{\mu\nu}=\chi_{I}^{}\,\frac{\mu_{B}^{}\,f_{\rho BN}}{2\,m_{\rho}}
\bar{u}(p^{\prime})\left[\Gamma_{VB}^{\mu\alpha}(k)\,S_{\alpha\beta}(p+q)\right.
\nonumber\\
&&\back \left.\quad \times\Gamma_{\gamma B}^{\nu\beta}(q)
+\Gamma_{\gamma B}^{\nu\alpha}(q)
\,S_{\alpha\beta}(p-k)\,\Gamma_{VB}^{\mu\beta}(k)\right]u(p),
\label{Mres}
\end{eqnarray}
for intermediate spin-1/2 and -3/2 states, respectively; $p$ and $p^{\prime}$
are the momenta of the in- and outgoing nucleon, $q$ ($k$) is the photon
($\rho$) momentum, and $\chi_{I}^{}=2$ $(4/3)$ an isospin factor for
$I=1/2$ $(3/2)$ resonances. The vertices $\Gamma$ follow from Eq.~(\ref{Lag}):
\begin{eqnarray}
&&\back \Gamma_{\gamma B}^{\mu}(q)=\Gamma_{VB}^{\mu}(q)=2\,\gamma_{5}\,\sigma^{\alpha\mu}\,q_{\alpha}\,F(\vec{q}\,) ,\\
&&\back \Gamma_{\gamma B}^{\mu\nu}(q)=\Gamma_{VB}^{\mu\nu}(q)=\left(\gamma_{5}\,\gamma^{\mu}\,q^{\nu}-\gamma_{5}\,\qslash\,g^{\mu\nu}\right)\,F(\vec{q}\,) ,
\nonumber
\end{eqnarray}
for positive parity resonances and
\begin{eqnarray}
&&\back \Gamma_{\gamma B}^{\mu}(q)=\Gamma_{VB}^{\mu}(q)=2\,\sigma^{\alpha\mu}\,q_{\alpha}\,F(\vec{q}\,) ,\\
&&\back \Gamma_{\gamma B}^{\mu\nu}(q)=\Gamma_{VB}^{\mu\nu}(q)=\left(\gamma^{\mu}\,q^{\nu}-\qslash\,g^{\mu\nu}\right)\,F(\vec{q}\,),
\nonumber
\end{eqnarray}
for negative parity resonances.

It is now straightforward to implement the production amplitude of
Ref.~\cite{Oh:2003aw}, augmented by baryon resonances, into a mass
differential cross section per nucleon for exclusive $\rho$ and 
dilepton production.  For the latter one obtains
\begin{eqnarray}
&&\back \left\langle\frac{d\sigma}{d\,M}\right\rangle_{A}(q,M)
=\frac{m_N^2\,M}{(2\pi)^2\,\rho_{A}^{}}
\nonumber\\
&&\back \times\,\int\frac{d^3p}{(2\pi)^3}\frac{d^4k}{(2\pi)^4}
\frac{d^{4}p^{\prime\,}}{(2\pi)^4}\frac{e^2\,g^2}{(k^2)^2\,m_\rho^4}
\frac{-\Im\,\Sigma_{\gamma\,\rightarrow\,e^+e^-}^{vac}(k)}
{2\,\sqrt{(p\cdot q)^2}}
\nonumber\\
&&\back \times\,\delta(k^2-M^2)\,\delta(p^{\prime\,2}-m_N^2)\,
\delta^4(q+p-k-p^\prime)
\nonumber\\
&&\back
\times\,\Theta\left(k_f-|\vec{p}\,| \right) \,
\sum\limits_{m_s,m_{s^\prime},\lambda}T^\mu(q,p,k)\,
\left(T^\nu(q,p,k)\right)^\dagger\,
\nonumber\\
&&\back \times\,
\left\{P^L_{\mu\nu}(k)\,|G_\rho^L(k)|^2+P^T_{\mu\nu}(k)\,|G_\rho^T(k)|^2\right\}\,,
\label{sigmaF}
\end{eqnarray}
where $T^\mu$ follows from summing $\mathcal{M}_i^{\mu\nu}$
over the photon polarization $\epsilon_\mu$ in the elementary
processes,
\begin{equation}
T^\mu(q,p,k)=\sum\limits_{i\in\left\{\sigma,\phi,N,f_2,B\right\}}\,
\mathcal{M}_i^{\mu\nu}(q,p,k)\,\epsilon_\nu(q)\, ,
\label{Tmu}
\end{equation}
and the dilepton final state is represented by
\begin{equation}
\back \Im\,\Sigma_{\gamma\,\rightarrow\,e^+e^-}^{vac}(k)
=-\frac{e^2\,k^2}{96\,\pi^2}\,.
\label{ImSigee}
\end{equation}
Eq.~(\ref{sigmaF}) contains an average over the Fermi motion of the
incoming nucleon ($p_0^2=\vec{p}^{\,2}+m_N^2$) as needed for nuclear
targets in the next section.
Furthermore, $G^{L/T}_\rho$ denote the longitudinal and transverse
components of the electromagnetic correlator~\cite{Rapp:1997ei} (in
naive VDM, one has
$G^{L/T}_\rho=(m_\rho^{(0)})^4/g_\rho^2 D_\rho^{L,T}$ where
$D_\rho^{L,T}$ is the $\rho$ propagator), and
$P^{L,T}_{\mu\nu}$ are projection operators. 
 
Our main interest in the present paper concerns the shape changes
in the dilepton mass spectra induced by the in-medium 
$\rho$ propagator encoded in $G_\rho^{L/T}$. Note, however, that 
Eq.~(\ref{sigmaF}) also accounts for the reduction in dilepton 
emission due to absorption of the $\rho$ meson propagating in a nuclear 
medium at fixed density, via the in-medium reduction of 
$|G_\rho^{L/T}|^2$. For finite nuclei, this effect causes an appreciable 
decrease of the total $e^+e^-$ production cross section, relative to a 
simple scaling with nuclear mass number, $A$ (referred to as nuclear 
transparency ratio, $T_A$; see, e.g., 
Refs.~\cite{Kaskulov:2006zc,Kotulla:2008xy,Ishikawa:2004id} for 
the cases of $\omega$ and $\phi$ photoproduction).

\begin{figure}[!t]
\includegraphics[scale=0.85]{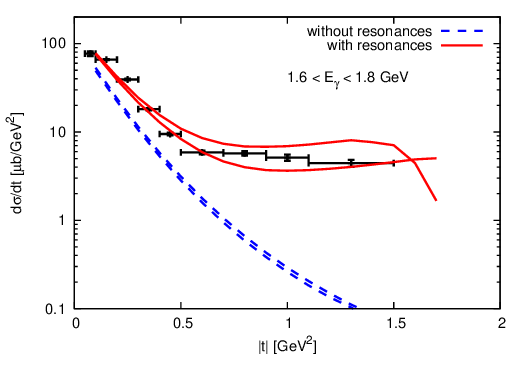}
\includegraphics[scale=0.85]{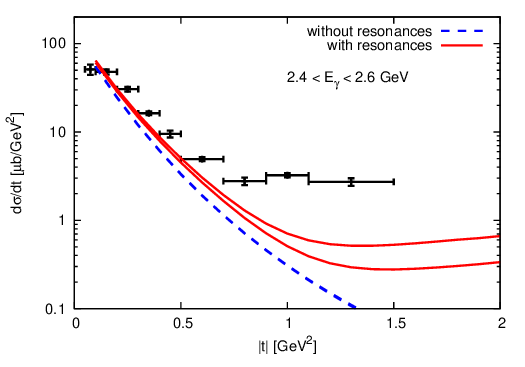}
\caption{Differential $\rho$-meson photoproduction cross
section. Calculations with (solid lines) and without (dashed lines)
resonance contributions are compared to data~\cite{Wu:2005wf}, with
2 curves each representing the upper and lower end of the experimental
photon energy window. In each case the full vacuum $\rho$-meson spectral
function has been used.}
\label{fig_dsigdt}
\end{figure}
We first test our production amplitude in the process
$\gamma p \to \rho^0 p$.
The photon-energy dependence of the
$\rho$-production cross section is shown in Fig.~\ref{fig_xsec-tot}.
The contribution of the resonances nicely fills in low-energy
strength that was missing in the original model of Ref.~\cite{Oh:2003aw}
and becomes negligible at energies beyond 2.5~GeV. Our calculations also
illustrate that the inclusion of the free $\rho$ width (as given by the
vacuum spectral function of Ref.~\cite{Rapp:1997ei}) further improves the
agreement with the low-energy cross section (however, in this region the
extraction of the data is beset with significant model
dependence~\cite{Wu:2005wf}). The scattering-angle differential
cross section (Fig.~\ref{fig_dsigdt}) reveals that the resonance
excitations provide large contributions at large scattering angle which
is supported by experiment.

\begin{figure}[!t]
\includegraphics[scale=0.85]{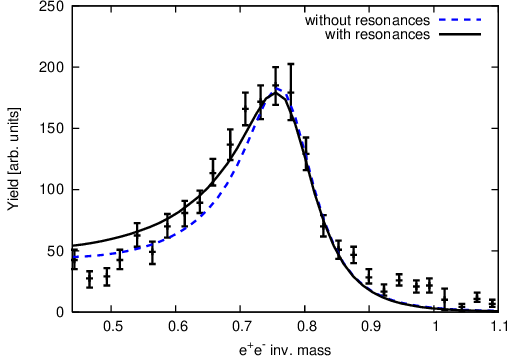}
\caption{Dilepton invariant-mass spectrum for $\rho$ photoproduction off
deuterium with (solid line) and without (dashed line) baryon resonance
contributions, compared to CLAS data after subtraction of $\omega$ and
$\phi$ contributions~\cite{Clas:2007mga,Wood:2008ee}.}
\label{fig_deut}
\end{figure}
Next we apply our model to dilepton invariant-mass spectra off deuterium.
To mimic a finite nucleon-momentum distribution and rescattering 
effects in $G_\rho$ we use a small average density of 0.1\,$\varrho_0$ in
Eq.~(\ref{sigmaF}) (folding over a realistic
density distribution gives similar results).
The incoming photon energies are weighted in 6 bins from $q_0=$~1-3.5\,GeV
to simulate the Bremsstrahlungs-spectrum used by CLAS~\cite{CLAS-priv}.
The shape of the $e^+e^-$ spectra~\cite{Wood:2008ee} is reasonably well 
reproduced, except for masses above 0.9\,GeV 
where additional production processes 
might become relevant, see also Ref.~\cite{Wood:2008ee}.

\section{Dilepton Spectra off Nuclei}
\label{sec_med}
\begin{figure}[!t]
\includegraphics[scale=0.85]{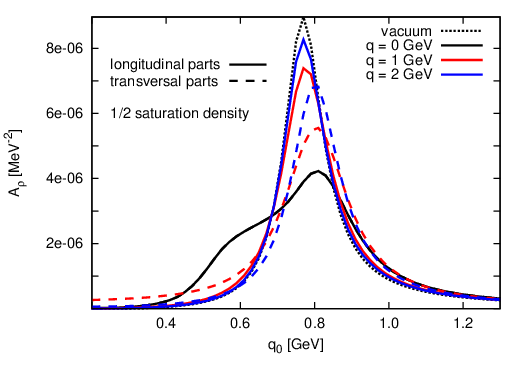}
\caption{In-medium $\rho$-meson spectral function at various 3-momenta
and nuclear density 0.5\,$\varrho_0$~\cite{Rapp:1999us};
solid (dashed) lines: transverse
(longitudinal) parts  (identical at $q=0$).}
\label{fig_Arho}
\end{figure}

To evaluate medium effects for nuclear targets, we first have to estimate
the densities probed for a given nucleus. If the $\rho$ meson were produced
at rest, the density at its creation point would be a good approximation.
However, since we are considering rather high photon energies the $\rho$
meson will travel a significant distance before it decays. 
Based on the assumption that the (medium effect on the) $\rho$ 
instantaneously adjusts to the surrounding medium, the relevant
density for the dilepton spectrum is the local density at the decay 
point, which we estimate as follows. For the incoming photon the 
interaction point is distributed according to a Woods-Saxon density 
profile (weighted by volume). The average travel distance of the 
$\rho$ from its production to decay point is then calculated as
\begin{eqnarray}
L=|\vec{v}|\,\gamma\,\tau , \qquad |\vec{v}|
=|\vec{k}\,|\left(\vec{k}^{\,2}+m_\rho^2\right)^{-1/2},
\end{eqnarray}
where $|\vec{v}|$ is the $\rho$ three-velocity and $\tau$ its average
lifetime (time dilated by a Lorentz $\gamma$ factor). Under the present
conditions, the latter is roughly $\sim 1$\,fm/c from the underlying
in-medium spectral function, cf.~Fig.~\ref{fig_Arho}.
The velocity is estimated from the incoming photon energy for an on-shell
$\rho$ in the limiting case of forward production where the bulk of the
differential cross section is concentrated (recall Fig.~\ref{fig_dsigdt},
where nuclear Fermi motion is neglected).
The travel length $L$ obtained in this way is then integrated over all
production points
resulting in the following distribution of decay points at a given
density $\varrho_x$,
\begin{eqnarray}
&&\back N(\varrho_x)=\!\int\!\varrho(r,0,\theta)\,r^2\,\sin(\theta)\,
\delta(\varrho(r,L,\theta)-\varrho_x)\,d^3r,
\nonumber\\
&&\back \varrho(r,L,\theta)=\frac{\varrho_0}
{1 + \exp\left[\frac{(r^2 + L^2 - 2\,r\,L\,\cos(\theta))^{1/2} - c}{z}\right]},
\end{eqnarray}
with $z=0.55$~fm and $c=4.05$~fm for iron.
At an average incoming photon energy of $\sim$2.1\,GeV (representative for
the CLAS experiment~\cite{CLAS-priv}) the average density at the decay point
amounts to $0.5\,\varrho_0$. Varying the photon energy between 1.5\,GeV 
and 2.5\,GeV affects the average density by about $\pm 0.1\,\varrho_0$. 
We therefore display our dilepton spectra on iron for a density range of 
$\varrho_N=$~0.4-0.6\,$\varrho_0$.\footnote{This also reflects 
some of the uncertainty introduced by letting all $\rho$ mesons decay 
after a fixed distance $L$, as compared to a distribution in $L$, 
since a different photon energy translates into a different $L$.}  
Note that lower photon energies (probing larger densities) imply a 
smaller $\rho$-meson phase space which is therefore biased toward lower
invariant masses. While the incoming photon energy spectrum is properly
included at a given density
via Eq.~(\ref{sigmaF}), the density-energy correlation is neglected. However,
across the above range, the density variation of the dilepton spectra turns
out to be quite moderate. A more accurate evaluation of this correlation
should also include an exponential decay distribution of the  $\rho$ decays
in $L$.

The main in-medium input to Eq.~(\ref{sigmaF}) is the $\rho$ spectral
function of Refs.~\cite{Rapp:1997ei,Rapp:1999us} which is displayed in
Fig.~\ref{fig_Arho} for transverse and longitudinal modes at various
3-momenta and at $\varrho_N^{}=0.5\,\varrho_0$. At 3-momenta relevant
for CLAS ($q \simeq 1\mbox{--2}$~GeV) the medium effects are
significantly reduced compared to $q=0$ (a consequence of the typical
formfactor cutoffs, $\Lambda_{\rho BN}\simeq 0.6$~GeV; the reduction is
more pronounced than, e.g., in the spectral function of
Ref.~\cite{Post:2003hu} due to larger formfactor cutoffs used in there).
In addition, a noticeable difference between longitudinal and transverse
modes develops, the latter exhibiting an upward mass shift which
is due to both pion cloud and $P$-wave resonance excitations.
Note that in applications to dilepton spectra at CERN-SPS the in-medium
spectral function is predominantly probed at 3-momenta below
1~GeV~\cite{Rapp:1999us,vanHees:2007th}. This reiterates the notion
that the CLAS data provide a novel test of the spectral function at high
3-momentum.

\begin{figure}[!t]
\includegraphics[scale=0.8]{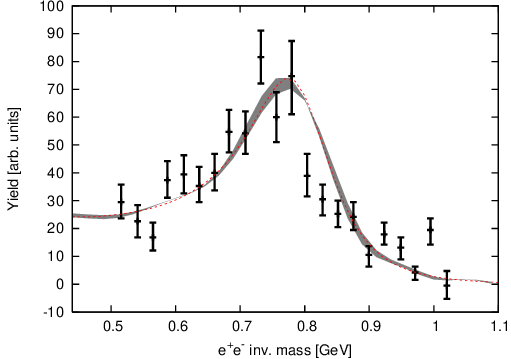}
\includegraphics[scale=0.8]{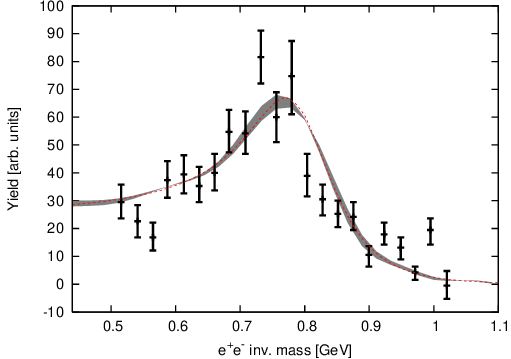}
\includegraphics[scale=0.8]{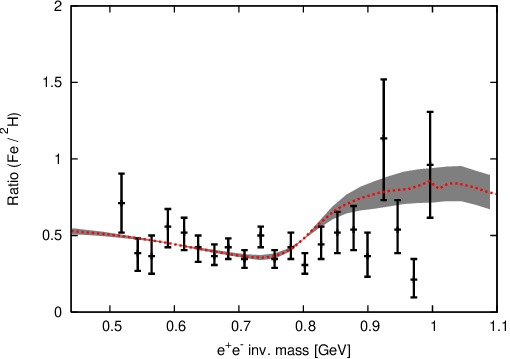}
\caption{Theoretical calculations of dilepton spectra for photoproduction
off iron compared to CLAS data~\cite{Clas:2007mga,Wood:2008ee}.
The bands represent the nuclear density range
$\varrho_N^{}=0.4\mbox{--}0.6~\varrho_0$. The curves in the
upper (middle) panel are calculated without (with) baryon resonances
(Tab.~\ref{tab_res}) in the elementary production amplitude, while the lower
panel shows the iron-to-deuteron ratio for the full calculation.}
\label{fig_iron}
\end{figure}

Our results for the dilepton invariant-mass spectra on iron are compared
to the CLAS ``excess" spectra in Fig.~\ref{fig_iron} using the density range
as estimated above. For each density, the normalization is adjusted
to the integrated strength of the data. Alternatively, one can determine
the normalization by a least-square fit
resulting in $\chi^2/N$=1.29 (1.4) per data point (not) including
the resonance contributions in the production process, compared to
$\chi^2/N$=1.34 (1.49) when normalizing to the data. 
In either case, the agreement with the data is fair
(the in-medium broadening of the nucleon
resonances has very little impact on the dilepton spectra).
A slight 
discrepancy with the data for masses of $M=0.8\mbox{--}0.85$~GeV may 
allow for a small attractive mass shift of about 
$-$15\,MeV.\footnote{In the transport-based~\cite{Muhlich:2002tu}
Breit-Wigner fits in Ref.~\cite{Wood:2008ee}, the extracted $\rho$-mass is 
consistent with the free mass.} 
Overall, the rather moderate medium effects in the (transverse and longitudinal
parts of the) $\rho$ spectral function at high 3-momentum (as seen in 
Fig.~\ref{fig_Arho}) are essentially in line with the CLAS spectra.\footnote{The reduction of $\Lambda_{\rho NN}$ to 600\,MeV entails an attraction 
of $\sim$15\,MeV at $\varrho_N^{}$=$\varrho_0$ in the transverse $\rho$ spectral
function.}
There are further effects which could modify our spectra at the several
percent level, e.g., in-medium $\omega$-meson decays along with
interference/mixing with the $\rho$. $\sigma$ and $f_2$ tadpole
diagrams are not included in the spectral function; implementing the
former with the coupling strength employed in the elementary production
process~\cite{Oh:2003aw} generates an attractive mass
shift of about $-10$~MeV for the iron target.  
Further processes in inclusive production (e.g., $\gamma+N\to\rho+N+\pi$), 
additional resonance strength to accommodate
large angle-scattering at high photon energies (recall lower panel in
Fig.~\ref{fig_dsigdt}), or a more elaborate treatment of the baryon-resonance
widths, might also play a role.

\section{Conclusions}
\label{sec_concl}
We have performed an essentially parameter-free calculation of $\rho$
photoproduction off nuclei, combining a realistic model for the
elementary production process with a hadronic many-body spectral
function~\cite{Rapp:1999us} which was extensively used before in the
interpretation of dilepton spectra in heavy-ion collisions. An earlier
constructed photoproduction amplitude~\cite{Oh:2003aw} has been
supplemented with resonance contributions as implicit in the
in-medium $\rho$ spectral function. A reasonable description of
$\rho$ photoproduction cross sections on the proton, as well as
dilepton spectra on deuterium, emerged without major adjustments.
The key test of the spectral function has been provided by the dilepton
(``excess") spectra off iron. With average densities estimated from
the decay kinematics for incoming photon energies as used at JLAB, the
rather moderate in-medium effects reported by the CLAS experiment are
fairly well reproduced.
The main difference compared to the stronger effects
observed in heavy-ion collisions is the rather large 3-momentum of the
$\rho$ in the CLAS data, for which the spectral function of
Ref.~\cite{Rapp:1999us} predicts a significantly reduced broadening.
Clearly, a low-momentum cut on the dilepton spectra would enable a
critical test of the predicted increase in medium effects.
Further constraints could be obtained by analyzing absolute $e^+e^-$
production cross sections (e.g., the so-called nuclear transparency ratio), 
as the in-medium spectral width of the $\rho$ is directly related to its 
absorption in the nuclear medium.

\vspace{0.5cm}

{\bf Acknowledgments}\\
The authors acknowledge useful discussions with C.~Djalali.
FR and RR were supported by a U.S. National Science Foundation 
CAREER grant No.~PHY-0449489.  
TSHL was supported by the U.S. Department of Energy,
Office of Nuclear Physics Division, under contract No.~DE-AC02-06CH11357.
YO was supported by the U.S. NSF under grants No.~PHY-0457265, 
PHY-0758155 and the Welch Foundation under Grant No.~A-1358.

\bibliography{../../Literature}
\bibliographystyle{prsty}


%

\end{document}